\documentclass[11pt]{article}
\usepackage{amsmath}
\usepackage{latexsym}
\usepackage{amssymb}
\usepackage{a4wide}
\usepackage{bbm}
\usepackage{epsfig}
\newcommand{\I}{\text{i}}
\newcommand{\E}{\text{e}}
\newcommand{\tr}{\text{tr}}
\newcommand{\Tr}{\text{Tr}}
\newcommand{\re}[1]{~(\ref{#1})}

\newcommand{\ts}{\tilde{s}}
\newcommand{\Fsf}{\mathsf{F}}
\begin{document}
\pagestyle{plain}

%
\newcommand{\be}{\begin{equation}}
\newcommand{\ee}{\end{equation}\noindent}
\newcommand{\bear}{\begin{eqnarray}}
\newcommand{\ear}{\end{eqnarray}\noindent}
\newcommand{\no}{\noindent}
\date{}
\renewcommand{\theequation}{\arabic{section}.\arabic{equation}}
\renewcommand{\arraystretch}{2.5}
\newcommand{\GeV}{\mbox{GeV}}
\newcommand{\cL}{\cal L}
\newcommand{\D}{\cal D}
\newcommand{\Dhalf}{{D\over 2}}
\newcommand{\Det}{{\rm Det}}
\newcommand{\PP}{\cal P}
\newcommand{\G}{{\cal G}}
\def\GBd12{{\dot G}_{B12}}
\def\R{1\!\!{\rm R}}
\def\Eins{\mathord{1\hskip -1.5pt
\vrule width .5pt height 7.75pt depth -.2pt \hskip -1.2pt
\vrule width 2.5pt height .3pt depth -.05pt \hskip 1.5pt}}
\newcommand{\symb}{\mbox{symb}}
\renewcommand{\arraystretch}{2.5}
\newcommand{\slD}{\raise.15ex\hbox{$/$}\kern-.57em\hbox{$D$}}
\newcommand{\slpartial}{\raise.15ex\hbox{$/$}\kern-.57em\hbox{$\partial$}}
\newcommand{\slG}{{{\dot G}\!\!\!\! \raise.15ex\hbox {/}}}
\newcommand{\Gd}{{\dot G}}
\newcommand{\Gund}{{\underline{\dot G}}}
\def\mn{{\mu\nu}}
\def\rs{{\rho\sigma}}
\def\np{n_{+}}
\def\nm{n_{-}}
\def\Np{N_{+}}
\def\Nm{N_{-}}
\def\PITD{{(4\pi T)}^{-{D\over 2}}}
\def\non{\nonumber}
\def\beqn*{\begin{eqnarray*}}
\def\eqn*{\end{eqnarray*}}
\def\sy{\scriptscriptstyle}
\def\footstrut{\baselineskip 12pt}
\def\square{\kern1pt\vbox{\hrule height 1.2pt\hbox{\vrule width 1.2pt
   \hskip 3pt\vbox{\vskip 6pt}\hskip 3pt\vrule width 0.6pt}
   \hrule height 0.6pt}\kern1pt}
\def\slash#1{#1\!\!\!\raise.15ex\hbox {/}}
\def\dint#1{\int\!\!\!\!\!\int\limits_{\!\!#1}}
\def\bra#1{\langle #1 |}
\def\ket#1{| #1 \rangle}
\def\vev#1{\langle #1 \rangle}
\def\rightvac{\mid 0\rangle}
\def\leftvac{\langle 0\mid}
\def\dps{\displaystyle}
\def\sy{\scriptscriptstyle}
\def\half{{1\over 2}}
\def\third{{1\over3}}
\def\fourth{{1\over4}}
\def\fifth{{1\over5}}
\def\sixth{{1\over6}}
\def\seventh{{1\over7}}
\def\eigth{{1\over8}}
\def\ninth{{1\over9}}
\def\tenth{{1\over10}}
\def\pa{\partial}
\def\ddtau{{d\over d\tau}}
\def\ge{\hbox{\textfont1=\tame $\gamma_1$}}
\def\gz{\hbox{\textfont1=\tame $\gamma_2$}}
\def\gd{\hbox{\textfont1=\tame $\gamma_3$}}
\def\go{\hbox{\textfont1=\tamt $\gamma_1$}}
\def\gt{\hbox{\textfont1=\tamt $\gamma_2$}}
\def\gth{\hbox{\textfont1=\tamt $\gamma_3$}} 
\def\gf{\hbox{$\gamma_5\;$}}
\def\ie{\hbox{$\textstyle{\int_1}$}}
\def\iz{\hbox{$\textstyle{\int_2}$}}
\def\id{\hbox{$\textstyle{\int_3}$}}
\def\ldop{\hbox{$\lbrace\mskip -4.5mu\mid$}}
\def\rdop{\hbox{$\mid\mskip -4.3mu\rbrace$}}
\def\eps{\epsilon}
\def\epshalf{{\epsilon\over 2}}
\def\e{\mbox{e}}
\def\g{\mbox{g}}
\def\pa{\partial}
\def\kinb{{1\over 4}\dot x^2}
\def\kinf{{1\over 2}\psi\dot\psi}
\def\expk{{\rm exp}\biggl[\,\sum_{i<j=1}^4 G_{Bij}k_i\cdot k_j\biggr]}
\def\expp{{\rm exp}\biggl[\,\sum_{i<j=1}^4 G_{Bij}p_i\cdot p_j\biggr]}
\def\expshort{{\e}^{\half G_{Bij}k_i\cdot k_j}}
\def\expabb{{\e}^{(\cdot )}}
\def\epseps#1#2{\varepsilon_{#1}\cdot \varepsilon_{#2}}
\def\epsk#1#2{\varepsilon_{#1}\cdot k_{#2}}
\def\kk#1#2{k_{#1}\cdot k_{#2}}
\def\G#1#2{G_{B#1#2}}
\def\Gp#1#2{{\dot G_{B#1#2}}}
\def\GF#1#2{G_{F#1#2}}
\def\Dab{{(x_a-x_b)}}
\def\Dsq{{({(x_a-x_b)}^2)}}
\def\lag{( -\partial^2 + V)}
\def\4piTD{{(4\pi T)}^{-{D\over 2}}}
\def\4piT4{{(4\pi T)}^{-2}}
\def\TintmD{{\dps\int_{0}^{\infty}}{dT\over T}\,e^{-m^2T}
    {(4\pi T)}^{-{D\over 2}}}
\def\Tintm4{{\dps\int_{0}^{\infty}}{dT\over T}\,e^{-m^2T}
    {(4\pi T)}^{-2}}
\def\Tintm{{\dps\int_{0}^{\infty}}{dT\over T}\,e^{-m^2T}}
\def\Tint{{\dps\int_{0}^{\infty}}{dT\over T}}
\def\pint{{\dps\int}{dp_i\over {(2\pi)}^d}}
\def\Dx{\dps\int{\cal D}x}
\def\Dy{\dps\int{\cal D}y}
\def\Dpsi{\dps\int{\cal D}\psi}
\def\Tr{{\rm Tr}\,}
\def\tr{{\rm tr}\,}
\def\sumij{\sum_{i<j}}
\def\freeexp{{\rm e}^{-\int_0^Td\tau {1\over 4}\dot x^2}}
\def\arraystretch{2.5}
\def\Ge{\mbox{GeV}}
\def\dA{\partial^2}
\def\DA{\sqsubset\!\!\!\!\sqsupset}
\def\FFdual{F\cdot\tilde F}
%
%
\def\bbbr{{\rm I\!R}}
\def\bbbone{{\mathchoice {\rm 1\mskip-4mu l} {\rm 1\mskip-4mu l}
{\rm 1\mskip-4.5mu l} {\rm 1\mskip-5mu l}}}
\def\bbbz{{\mathchoice {\hbox{$\sf\textstyle Z\kern-0.4em Z$}}
{\hbox{$\sf\textstyle Z\kern-0.4em Z$}}
{\hbox{$\sf\scriptstyle Z\kern-0.3em Z$}}
{\hbox{$\sf\scriptscriptstyle Z\kern-0.2em Z$}}}}
\pagestyle{empty}
\renewcommand{\thefootnote}{\fnsymbol{footnote}}
\hskip 9cm {\sl CERN-TH/2001-097}
\vskip-.1pt
\hskip 9cm {\sl UMSNH-Phys/01-4}
\vskip .4cm
\begin{center}
{\Large\bf Vacuum Polarisation Tensors in}\\ 
{\Large\bf Constant Electromagnetic Fields: Part III}
\vskip1.3cm

{\large Holger Gies}
\\[1.5ex]
{\it
Theory Division, CERN\\
CH-1211 Geneva 23, Switzerland\\
Holger.Gies@cern.ch\\
}
\vskip.5cm

\vskip.5cm
 {\large Christian Schubert
}
\\[1.5ex]
{\it
Instituto de F\'{\i}sica y Matem\'aticas
\\
Universidad Michoacana de San Nicol\'as de Hidalgo\\ 
Apdo. Postal 2-82\\
C.P. 58040, Morelia, Michoac\'an, M\'exico\\
schubert@itzel.ifm.umich.mx\\
}
and\\
{\it
California Institute for Physics and Astrophysics\\
366 Cambridge Ave., Palo Alto, California, US}
\vskip 1.5cm
\vspace{.5cm}

\vskip 2.5cm
 {\large \bf Abstract}
\end{center}
\begin{quotation}
\noindent
The string-inspired technique is used for
a first calculation of the one-loop
axialvector vacuum polarisation
in a general constant electromagnetic field.
A compact result is reached for the difference
between this tensor and the corresponding 
vector vacuum polarisation. This result is confirmed
by a Feynman diagram calculation. Its
physical relevance is briefly discussed.

\end{quotation}
\clearpage
\renewcommand{\thefootnote}{\protect\arabic{footnote}}
\pagestyle{plain}

\setcounter{page}{1}
\setcounter{footnote}{0}

\section{Introduction: Standard Model
Processes in Constant Electromagnetic Fields}
\renewcommand{\theequation}{1.\arabic{equation}}
\setcounter{equation}{0}

Following the calculation of the one-loop 
vector--vector and vector--axialvector vacuum
polarisation tensors in a general constant
electromagnetic field, presented in parts
I \cite{vv} and II \cite{va} of this series,
in the present third part we consider the
axialvector--axialvector case. 
As in the previous cases, we will use the
``string-inspired'' worldline path integral
formalism 
\cite{berkos}--\cite{review} 
to arrive at a compact integral
representation of this quantity. 
As a check we will also perform a Feynman
diagrammatic calculation of it.
With both methods it will turn out to 
be considerably simpler not to compute the axialvector 
vacuum polarisation itself, but rather the difference
between this and
the known \cite{batsha,urrutia,Dittrich:2000zu,va}
vector vacuum polarisation in a constant field.

As in the case of the vector--axialvector amplitude, considered in
part II, our main physical interest in this quantity stems from its
relevance for low-energy neutrino processes. In particular, we refer
to processes where the external momentum flux through heavy
gauge boson propagators remains small compared to $m_W$, 
neutrino energies $E_\nu\ll m_W^3/eF$ and field strengths $eF\ll
m_W^2$, so that the local limit of the standard model interaction,
the Fermi theory, is applicable ($m_W$ denotes the heavy gauge boson
mass). Vacuum polarisation, or phrased differently, the virtual
existence of the neutrinos as charged lepton pairs, transfers
electromagnetic properties to the neutrinos without requiring
additional non-standard parameters (magnetic moments etc.). These
loop-induced properties allow for neutrino--photon interactions or
interactions of neutrinos with external electromagnetic fields.

The amplitude considered here occurs, for instance, in scattering
processes involving 4 neutrinos and an arbitrary number of soft
photons, and in decay processes involving 2 neutrinos and a lepton
pair in an external field. For magnetic fields, the latter have been
studied intensively in
\cite{botezh}--\cite{Hardy:2001gg}, 
and it is believed that processes of this type can contribute
significantly to neutrino energy loss in astrophysical processes
involving extreme conditions. Therefore, neutrino heating and cooling
processes can be partly governed by those neutrino interactions
enhanced by electromagnetic fields. The present work allows for a
generalization of such results for magnetic fields to the case of a
general electromagnetic field; this provides for new dimensions in
parameter space involving electromagnetic invariants also with
electric components.  Although $E\ll B$ in most realistic scenarios,
the invariant ${\cal G}=\mathbf{E\cdot B}$ can have a sizeable value
and, moreover, owing to its pseudoscalar nature, allow for processes
that are forbidden in a purely magnetic field (see, e.g.,
\cite{Gies:2000wc}).

\section{Worldline Calculation of the Axialvector
\\ Vacuum Polarisation Tensor in a Constant Field} 
\renewcommand{\theequation}{2.\arabic{equation}}
\setcounter{equation}{0}

According to the formalism developed in \cite{mcksch,dimcsc,va} 
the one-loop
axialvector vacuum polarisation tensor in a constant field
can be represented as the following integral of
a worldline correlator of two axialvector
vertex operators
\footnote{%
We work initially in the Euclidean spacetime with
a positive definite metric
$g_{\mu\nu}={\,\mathrm{diag}}(++++)$.
The Euclidean field strength tensor is defined by
$F^{ij}= \varepsilon_{ijk}B_k, i,j = 1,2,3$,
$F^{4i}=-iE_i$, its dual by
$\tilde F^{\mu\nu} = \half 
\varepsilon^{\mu\nu\alpha\beta}F^{\alpha\beta}$
with $\varepsilon^{1234} = 1$.  
The corresponding Minkowski space amplitudes
are obtained by rotating 
$g_{\mu\nu}\rightarrow \eta_{\mu\nu}
= {\,\mathrm{diag}}(-+++)$, $
k^4\rightarrow -ik^0, T\rightarrow is,
\varepsilon^{1234}
\rightarrow
i\varepsilon^{1230},
\varepsilon^{0123}=1,
F^{4i}\rightarrow F^{0i}=E_i,
\tilde F^{\mu\nu}\rightarrow -i\tilde F^{\mu\nu}$.
}:

\bear
\Pi_{55}^\mn (k) &=&
e_5^2 \langle
A^{\mu}_5(k)A^{\nu}_5(-k)
\rangle,
\non\\
\langle
A^{\mu}_5(k_1)A^{\nu}_5(k_2)
\rangle
&=&
2\int_0^{\infty}
{dT\over T}
\e^{-m^2T}
(4\pi T)^{-{D\over 2}}
{\rm det}^{-{1\over 2}}
\biggl[{\tan({\cal Z})\over {{\cal Z} }}
\biggr] 
\int_0^Td\tau_1
\int_0^Td\tau_2
\nonumber\\
&&\hspace{1pt}\times
\biggl\langle
\Bigl(
ik_1^{\mu}+2\psi^{\mu}(\tau_1)\dot x(\tau_1)\cdot\psi(\tau_1)
+ \sqrt{D-2}\, z^{\mu}(\tau_1)
\Bigr)
\,\e^{ik_1\cdot x(\tau_1)}
\non\\
&&\hspace{1pt}\times
\Bigl(
ik_2^{\nu}+2\psi^{\nu}(\tau_2)\dot x(\tau_2)\cdot\psi(\tau_2)
+ \sqrt{D-2}\, z^{\nu}(\tau_2)
\Bigr)
\,\e^{ik_2\cdot x(\tau_2)}
\biggr\rangle.
\non\\
\label{P551}
\ear
Here $T$ denotes the global Schwinger proper-time variable for
the loop fermion, 
and ${\cal Z}_{\mu\nu} = eF_{\mu\nu}T$ with 
$F_{\mu\nu}$ the constant field strength tensor.
The spacetime dimension $D$ enters through
dimensional regularisation.
On the right-hand side
the angular brackets denote Wick contraction using the basic
field-dependent worldline correlators:

\begin{eqnarray}
\langle y^{\mu}(\tau_1)y^{\nu}(\tau_2)\rangle
&=&
-{\cal G}_B^{\mu\nu}(\tau_1,\tau_2)
= -
{T\over 2}
\biggl[{1\over {({\cal Z})}^2}
\biggl({{\cal Z}\over{{\rm sin}({\cal Z})}}
{\rm e}^{-i{\cal Z}\dot G_{B12}}
+i{\cal Z}\dot G_{B12} -1\biggr)
\biggr]^{\mu\nu},
\nonumber\\
\langle\dot y^{\mu}(\tau_1)y^{\nu}(\tau_2)\rangle
&=&
-\dot{\cal G}_B^{\mu\nu}(\tau_1,\tau_2)
=
-
\biggl[
{i\over {\cal Z}}\biggl({{\cal Z}\over{{\rm sin}({\cal Z})}}
{\rm e}^{-i{\cal Z}\dot G_{B12}}-1\biggr)
\biggr]^{\mu\nu},
\nonumber\\
\langle \dot y^{\mu}(\tau_1)\dot y^{\nu}(\tau_2)\rangle
&=&
\ddot {\cal G}_B^{\mu\nu}(\tau_1,\tau_2)
\,\,\,\,
=
2\delta(\tau_1-\tau_2) g^{\mu\nu}
-{2\over T}
\biggl[{{\cal Z}\over{{\rm sin}({\cal Z})}}
{\rm e}^{-i{\cal Z}\dot G_{B12}}
\biggr]^{\mu\nu},
\nonumber\\
\langle\psi^{\mu}(\tau_1)\psi^{\nu}(\tau_2)\rangle
&=&
\frac{1}{2}{\cal G}_F^{\mu\nu}(\tau_1,\tau_2)
\,= \frac{1}{2}
G_{F12}
\biggl[{{\rm e}^{-i{\cal Z}\dot G_{B12}}\over {\rm cos}({\cal Z})}
\biggr]^{\mu\nu},
\label{exfieldgreens}
\end{eqnarray}
where 
\bear
\dot G_{B12} &=& {\rm sign}(\tau_1 - \tau_2)
- 2 {{(\tau_1 - \tau_2)}\over T},\nonumber\\
G_{F12}&=&
{\rm sign}(\tau_1-\tau_2),
\label{defGBpGF}
\ear
and the trigonometric expressions should
be understood as power series in the Lorentz matrix
${\cal Z}$.
The field $z$ is auxiliary and has a
trivial correlator:
\bear
\langle z^{\mu}(\tau_1)z^{\nu}(\tau_2)\rangle
&=&
2\delta (\tau_1-\tau_2)g^{\mu\nu}.
\label{wickz}
\ear\no
After explicit Wick contraction, the expression in
angular brackets becomes
\bear
\langle \cdots \rangle_{A_5A_5}
&=&
\e^{-k\cdot \bar {\cal G}_{B12}\cdot k}
\Biggl\lbrace
k^{\mu}k^{\nu}
+k^{\mu}\Bigl[{\cal G}_{F22}
(\dot {\cal G}_{B21}-\dot{\cal G}_{B22})
k\Bigr]^{\nu}
+
k^{\nu}\Bigl[{\cal G}_{F11}
(\dot {\cal G}_{B12}-\dot{\cal G}_{B11})
k\Bigr]^{\mu}
\nonumber\\
&&\hspace{-70pt}
+
\biggl(
{\cal G}_{F11}^{\mu\rho}
{\cal G}_{F22}^{\nu\sigma}
-
{\cal G}_{F12}^{\mu\nu}
{\cal G}_{F12}^{\rho\sigma}
+
{\cal G}_{F12}^{\mu\sigma}
{\cal G}_{F12}^{\rho\nu}
\biggr)
\biggl(
\ddot {\cal G}_{B12}^\rs
-
\Bigl[(\dot {\cal G}_{B11}-\dot {\cal G}_{B12})k\Bigr]^{\rho}
\Bigl[(\dot {\cal G}_{B21}-\dot {\cal G}_{B22})k\Bigr]^{\sigma}
\biggr)
\non\\&&
+2(D-2)\delta_{12}\,g^\mn
\Biggr\rbrace,
\non\\
\label{bigwick}
\ear
where $k=k_1=-k_2$ and 

$$\bar{\cal G}_{B12}\equiv {\cal G}_{B}
(\tau_1,\tau_2) -{\cal G}_{B}(\tau,\tau)
=
{T\over 2{\cal Z}}
\biggl(
{\e^{-i\dot G_{B12}{\cal Z}}-\cos({\cal Z})\over\sin({\cal Z})}
+i\dot G_{B12}\biggr).
$$

We write out the integrand explicitly using 
(\ref{exfieldgreens})
\footnote{We remark that care must be taken in the
determination of coincidence limits, owing to the
sign--function appearing in $\dot G_B$. For
example, 
$\lim_{\tau_2\rightarrow\tau_1}{\cal G}_{F12}^\mn
= -i(\tan ({\cal Z}))^\mn$, but
$\lim_{\tau_2\rightarrow\tau_1}
{\cal G}_{F12}^\mn {\cal G}_{F12}^{\rho\sigma}
= 
g^\mn g^\rs -(\tan({\cal Z}))^\mn(\tan({\cal Z}))^\rs
\ne 
{\cal G}_{F11}^\mn {\cal G}_{F11}^{\rho\sigma}
$.}.
Some terms involve products of Lorentz
matrices and can be simplified, for instance

\bear
{\cal G}_{F12}^\rs
\Bigl[(\dot {\cal G}_{B12}-\dot{\cal G}_{B11})k\Bigr]^{\rho}
\Bigl[(\dot {\cal G}_{B21}-\dot{\cal G}_{B22})k\Bigr]^{\sigma}
&=&
G_{F12}\,k\cdot 
\biggl[
{1+\cos^2({\cal Z})\over \cos({\cal Z})\sin^2({\cal Z})}
\cos(\dot G_{B12}{\cal Z})
\non\\&&\hspace{70pt}
-{2\over\sin^2({\cal Z})}
\biggr]\cdot k.
\non\\
\label{exampmult}
\ear
The result reads

\bear
\langle \cdots \rangle_{A_5A_5} &=&
\e^{-k\cdot \bar {\cal G}_{B12}\cdot k}
\Biggl\lbrace
-
\biggl[
\Bigl({{\cal E}_{12}\over\sin({\cal Z})}
-{1\over\sin({\cal Z})\cos({\cal Z})}
\Bigr)
k\biggr]^{\mu}
\biggl[
\Bigl({{\cal E}_{21}\over\sin({\cal Z})}
-{1\over\sin({\cal Z})\cos({\cal Z})}
\Bigr)
k\biggr]^{\nu}
\non\\
&&\hspace{-50pt} 
+
\Bigl({{\cal E}_{12}\over\cos({\cal Z})}k\Bigr)^{\mu}
\Bigl({{\cal E}_{21}\over\cos({\cal Z})}k\Bigr)^{\nu}
-\Bigl({{\cal E}_{12}\over\cos({\cal Z})}\Bigr)^{\mu\nu}
k\cdot\biggl[
{1+\cos^2({\cal Z})\over \cos({\cal Z})\sin^2({\cal Z})}
\cos(\dot G_{B12}{\cal Z})
-{2\over\sin^2({\cal Z})}
\biggr]\cdot k
\non\\
&&\hspace{-50pt} 
+{2\over T}
\biggl[
\Bigl({{\cal E}_{12}\over\cos({\cal Z})}\Bigr)^{\mu\nu}
{\rm tr}\Bigl({{\cal Z}\over\sin({\cal Z})\cos({\cal Z})}\Bigr)
-\Bigl({(1+\sin^2({\cal Z})){\cal Z}{\cal E}_{12}\over
\sin({\cal Z})\cos^2({\cal Z})}\Bigr)^{\mu\nu}
\biggr]
-2\delta_{12}g^\mn
\Biggr\rbrace,
\label{bigwick2}
\ear
where we abbreviated $\e^{-i\dot G_{B12} {\cal Z}}\equiv {\cal E}_{12}$.
Since from the axial Ward identity we know that, in the massless
case,
the axialvector--axialvector amplitude must coincide with the
vector--vector amplitude, we subtract from this the corresponding
integrand for the vector--vector case, given in
section 4 of part I:

\bear
\langle \cdots \rangle_{AA}
&=&
\biggl\langle
\Bigl(
\dot x^{\mu}(\tau_1)+2i\psi^{\mu}(\tau_1)
k_1\cdot\psi(\tau_1)
\Bigr)
\e^{ik_1\cdot x_1}
\Bigl(
\dot x^{\nu}(\tau_2)+2i\psi^{\nu}(\tau_2)
k_2\cdot\psi(\tau_2)   
\Bigr)
\e^{ik_2\cdot x_2}
\biggr\rangle
\non\\&=&
\e^{-k\cdot \bar {\cal G}_{B12}\cdot k}
\Biggl\lbrace
\ddot{\cal G}^{\mu\nu}_{B12}
-{\cal G}^{\mu\nu}_{F12}k\cdot{\cal G}_{F12}\cdot k
\non\\&&\hspace{-15pt}
- \biggl[
\Bigl(\dot{\cal G}_{B11}
-{\cal G}_{F11}
-\dot{\cal G}_{B12}\Bigr)
^{\mu\lambda}
  \Bigl(\dot{\cal G}_{B21}
-\dot{\cal G}_{B22}
+{\cal G}_{F22}
\Bigr)^{\nu\kappa}
+
{\cal G}^{\mu\lambda}_{F12}
{\cal G}^{\nu\kappa}_{F21}
\biggr]
k^{\kappa}k^{\lambda}
\Biggr\rbrace.
\non\\
\label{wickvv}
\ear\no
This indeed leads to some simplification,

\bear
\langle \cdots \rangle_{A_5A_5}
- \langle \cdots\rangle_{AA}
&=&
\e^{-k\cdot \bar {\cal G}_{B12}\cdot k}
\Biggl\lbrace
-4\delta_{12}g^\mn
+2\Bigl({{\cal E}_{12}\over\cos({\cal Z})}\Bigr)^{\mu\nu}
k\cdot
{\cal U}_{12}
\cdot k
\non\\
&&\hspace{-30pt}
+{2\over T}
\biggl[
\Bigl({{\cal E}_{12}\over\cos({\cal Z})}\Bigr)^{\mu\nu}
{\rm tr}\Bigl({{\cal Z}\over\sin({\cal Z})\cos({\cal Z})}\Bigr)
-2\Bigl({\sin({\cal Z}){\cal Z}{\cal E}_{12}\over
\cos^2({\cal Z})}\Bigr)^{\mu\nu}
\biggr]
\Biggr\rbrace,
\non\\
\label{diffaavv}
\ear
where 

$${\cal U}_{12}\equiv {1-\cos({\cal Z}\dot G_{B12})\cos({\cal Z})
\over \sin^2({\cal Z})}$$

\noindent
was introduced in part II, eq. (4.9).
The difference should vanish in the massless case, i.e. the
integrand should turn into a total derivative. And indeed,
if one adds to the above the following two total derivative terms
\footnote{At this point 
the reader should be warned that a naive application of
the chain rule to expressions involving sign$(\tau)$ can lead to
errors. For example, for $n$ odd one has
${\partial\over\partial\tau_1}(\dot G_{B12})^n
= 2\delta_{12} -{2n\over T}(\dot G_{B12})^{n-1}
\ne n\ddot G_{B12}(\dot G_{B12})^{n-1}
$.}, 

\bear
0 &=& {8\over (4\pi)^{D\over 2}}
\int_0^{\infty}dT{\partial\over\partial T}
\biggl\lbrace
{\e^{-m^2T}\over T^{1+{D\over 2}}}
{\rm det}^{-{1\over 2}}
\biggl[{\tan({\cal Z})\over {{\cal Z} }}
\biggr] 
\int_0^Td\tau_1
\int_0^Td\tau_2
\,\e^{-k\cdot \bar {\cal G}_{B12}\cdot k}
\Bigl({{\cal E}_{12}\over\cos({\cal Z})}\Bigr)^\mn
\non\\&&\hspace{80pt}
-T^{1-{D\over 2}}g^\mn +T^{2-{D\over 2}}
\Bigl[m^2+{k^2\over 6}]g^\mn
\biggr\rbrace\non\\
&&
+{4\over (4\pi)^{D\over 2}}
\int_0^{\infty}
{dT\over T^{1+{D\over 2}}}
\e^{-m^2T}
{\rm det}^{-{1\over 2}}
\biggl[{\tan({\cal Z})\over {{\cal Z} }}
\biggr] 
\int_0^Td\tau_1\int_0^Td\tau_2
\non\\&&\hspace{80pt}\times
{\partial\over\partial\tau_1}
\biggl\lbrace
\e^{-k\cdot \bar {\cal G}_{B12}\cdot k}
\dot G_{B12}
\Bigl({{\cal E}_{12}\over\cos({\cal Z})}\Bigr)^\mn
\biggr\rbrace,
\non\\
\label{totalderivative}
\ear
a cancellation of terms ensues, which is complete
in the massless case
\footnote{Note that we do not need to put
$D=4$ for this cancellation mechanism to work.
This confirms that, as stated in \cite{dimcsc},
the path integral construction given in \cite{mcksch}
does not break the chiral symmetry for parity-even loops.}.
In the massive case a single term survives,
leading to

\bear
\langle
A^{\mu}_5(k)A^{\nu}_5(-k)
\rangle
-
\langle A^{\mu}(k)A^{\nu}(-k)\rangle
&=&
-{8m^2\over (4\pi)^{D\over 2}}
\int_0^{\infty}
{dT \over T^{1+{D\over 2}}}
\e^{-m^2T}
{\rm det}^{-{1\over 2}}
\biggl[{\tan({\cal Z})\over {{\cal Z} }}
\biggr] 
\non\\&&\times
\int_0^T  d\tau_1 d\tau_2
\biggl[
{{\cal E}_{12}
\over\cos ({\cal Z})}
\biggr]^\mn
\e^{-k\cdot\bar {\cal G}_{B12}\cdot k}
\non\\
&&\hspace{-170pt}
=
-{8 m^2\over (4\pi)^{D\over 2}}
\int_0^{\infty}
{dT\over T}
T^{2-{D\over 2}}
\e^{-m^2T}
{\rm det}^{-{1\over 2}}
\biggl[{\tan({\cal Z})\over {{\cal Z} }}
\biggr] 
\int_0^1 du_1
\biggl[
{\cos(\dot G_{B12}{\cal Z})
\over\cos ({\cal Z})}
\biggr]^\mn
\e^{-Tk\cdot\Phi_{12}\cdot k}.
\non\\
\label{AAminusVV}
\ear
Here as usual we have rescaled to the unit circle,
$\tau_{1,2}=Tu_{1,2}$, 
$k\cdot \bar {\cal G}_{B12} \cdot k
\equiv Tk\cdot\Phi_{12}\cdot k$,
and set $u_2=0$. Only the cosine part of
${\cal E}_{12}$ contributes to the integral.

This integral still contains a logarithmic ultraviolet
divergence at $T=0$,
which becomes obvious if one sets the external field
equal to zero:

\begin{equation}
\langle
A^{\mu}_5(k)A^{\nu}_5(-k)
\rangle 
-
\langle A^{\mu}(k)A^{\nu}(-k)\rangle
{\stackrel{F=0}{=}}
-\frac{m^2}{2\pi^2} 
\int_0^{\infty}{dT\over T^{{D\over 2}-1}}
\int_0^1 du_1
\,\e^{-T(m^2+\frac{1}{4}(1-\dot G_{B12}^2)k^2)}
g^{\mu\nu}. \label{N4}
\end{equation}

Similar to the renormalisation of the vector vacuum
polarisation tensor in part I, we remove this divergence by
subtracting the same expression at vanishing field and momentum; the
meaning of this subtraction will be discussed below.  In this way we
obtain for the renormalised axialvector vacuum polarisation tensor

\bear
\bar\Pi^\mn_{55}(k)\!
&=&\!
{e_5^2\over e^2}\bar\Pi^\mn_{\rm spin}(k)
+\bar N^\mn_{55}(k),
\non\\
\bar N^\mn_{55}(k)\! &=&\!
-{e_5^2 m^2\over 2\pi^2}\!
\int_0^{\infty}
{dT\over T}
\e^{-m^2T}
\biggl\lbrace
{\rm det}^{-{1\over 2}}
\biggl[{\tan({\cal Z})\over {{\cal Z} }}
\biggr] 
\int_0^1\! du_1
\biggl[
{\cos(\dot G_{B12}{\cal Z})
\over\cos ({\cal Z})}
\biggr]^\mn\!
\e^{-Tk\cdot\Phi_{12}\cdot k}
-g^\mn\biggl\rbrace,
\non\\
\label{P55renorm}
\ear with $\bar\Pi^\mn_{\rm spin}$ as given in part I, Eqs.~(4.9) and
(4.11); here, the overbar characterises renormalised quantities.
Using the decomposition formulas from section 3.2 in part I, and
continuing to Minkowski space, we obtain our final result for this
amplitude:

\bear
\bar\Pi^\mn_{55}(k)
&=&
{e_5^2\over e^2}\bar\Pi^\mn_{\rm spin}
+{e_5^2 m^2\over 2\pi^2}
{\dps\int_{0}^{\infty}}{ds\over s}
\,e^{-ism^2}
\int_{-1}^1 {dv\over 2}
\Biggl\lbrace
{z_+z_-\over \tanh(z_+)\tanh(z_-)}
\non\\&&\times
{\rm exp}\biggl[
-is\sum_{\alpha=+,-}
{\cosh(z_{\alpha}v)-\cosh(z_{\alpha})
\over 2z_{\alpha}\sinh(z_{\alpha})}\,
k\cdot \hat{\cal Z}_{\alpha}^2\cdot k
\biggr]
\sum_{\alpha =+,-}
{\cosh(z_{\alpha}v)\over\cosh(z_{\alpha})}
\bigl(\hat{\cal Z}_{\alpha}^2\bigr)^{\mu\nu}
+\eta^\mn
\Biggr\rbrace,
\non\\
\label{vp55final}
\ear
where $v=\dot G_{B12}=1-2u_1$,
\bear
z_+ &=& iesa, \non\\
z_- &=& -esb, \non\\
({\hat{\cal Z}_+}^2)^\mn &=& {(F^2)^\mn-b^2\eta^\mn\over a^2+b^2},
\non\\
({\hat{\cal Z}_-}^2)^\mn &=& -{(F^2)^\mn+a^2\eta^\mn\over a^2+b^2},
\non\\
\label{minkdefs}
\ear
and $a,b$ are the secular invariants:
\bear
a &=& \sqrt{\sqrt{{\cal F}^2+{\cal G}^2}+{\cal F}},
\non\\
b &=& \sqrt{\sqrt{{\cal F}^2+{\cal G}^2}-{\cal F}},
\label{secular}
\ear
with

\bear
{\cal F} &=& \half (B^2-E^2), \non\\
{\cal G} &=&
{\bf E}\cdot{\bf B}.
\non\\
\label{defMaxinv}
\ear

We have verified that this integral representation agrees for $b\to 0$
with the result for the magnetic special case given in \cite{kuzmik}.
The field-free case has, of course, been known for a long time (see,
e.g., \cite{ahkkm}).  Note that the new contribution is not
transversal, i.e. that $k_\mu \bar N_{55}^{\mu\nu}\neq 0$; this is not
astonishing, since there is no Ward identity that could protect the
transversality of $\Pi_{55}^{\mu\nu}$.

It remains to elucidate the significance of the counterterm introduced
above. Its meaning depends on the context. Since it is proportional to
$A_5^2$, it obviously corresponds to the introduction of a mass term
for the axial gauge field, indicating that the ``axial QED''
consisting only of a massive fermion coupled to an axial gauge field
is not renormalisable. 

In the context of Fermi theory,
on the other hand, the axialvector-field coupling is associated with
the coupling of a left-handed neutrino current to the electron loop,
and $e_5$ is proportional to $g_{\text{A}} G_{\text{F}}$, i.e. the
axial coupling to the electron times Fermi's constant. Subtracting the
divergence mentioned above thereby corresponds to a
``renormalisation'' of Fermi's constant (disregarding the fact that
Fermi's theory is generally non-renormalisable). 

Finally, in the
context of the renormalisable microscopic theory, the Standard Model,
the present divergence has two origins; in order to recognize this,
consider the diagrams in Fig.~1, which, among others, contribute to our
amplitude.
\begin{figure}[t]
\begin{center}
\begin{picture}(130,30)
\put(-30,-260){\epsfig{figure=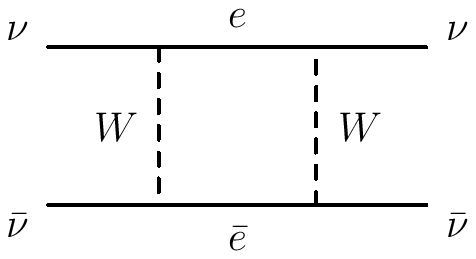,width=23cm}}
\put(50,-260){\epsfig{figure=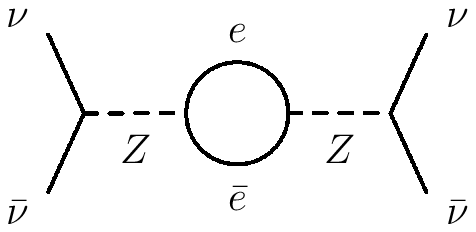,width=23cm}}
\end{picture}
\end{center}
\caption{$W$-boson (left-hand side) and $Z$-boson (right-hand side)
  exchange contributing to the axialvector amplitude in the context of
  Standard-Model neutrino interactions.}
\end{figure}
The left-hand side depicts a $W$-boson exchange diagram that is
finite by simple power-counting owing to the $1/p^2$ momentum
dependence of the $W$-boson propagator for large loop momentum $p^2\gg
m_W^2$. But upon taking the local limit by approximating the
propagator by $1/m_W^2$, in order to arrive at the effective Fermi
theory, artificial divergences are introduced. By contrast, the
right-hand side of Fig.~1 depicts a $Z$-boson exchange with a
self-energy correction to the boson. This diagram is, of course,
divergent before taking the local limit, and this divergence is
absorbed in a wave function and coupling constant renormalisation of
the $Z$-boson. The subtraction described above takes both types of
divergences into account and is normalised by requiring that vacuum
polarization has no physical effect on the axialvector field at zero
momentum if there are no external fields.

%

\section{Feynman Diagram Calculation of the Axialvector\\
Vacuum Polarisation Tensor in a Constant Field}
\renewcommand{\theequation}{3.\arabic{equation}}
\setcounter{equation}{0}

As a check of this result, we will now perform
a standard Feynman diagram calculation of this
quantity (in Minkowski space).

\subsection*{Decomposition Formulas}

We will use the same approach as before to the
decomposition of the field strength tensor,
although in the slightly different conventions
of \cite{bakast,shaisultanov2000,giesha}.
Defining

\begin{equation}
C_{\mu\nu}=\frac{1}{a^2+b^2} (-b\, F_{\mu\nu} +a\, \tilde{F}_{\mu\nu}), 
\quad
B_{\mu\nu}=\frac{1}{a^2+b^2} (a\, F_{\mu\nu} +b\, \tilde{F}_{\mu\nu}),
\label{CB}
\end{equation}
we have the relations

\begin{equation}
C_{\mu\nu}^2=\frac{1}{a^2+b^2} (F_{\mu\nu}^2 +a^2\, \eta_{\mu\nu}), \quad
B_{\mu\nu}^2=\frac{1}{a^2+b^2} (F_{\mu\nu}^2 -b^2\,
\eta_{\mu\nu}). \label{CqBq}
\end{equation}
The inverse relations are easily found:
\begin{eqnarray}
&&F_{\mu\nu}=-b\, C_{\mu\nu} +a\,B_{\mu\nu}, \quad
\tilde{F}_{\mu\nu}=a\, C_{\mu\nu} +b\, B_{\mu\nu}, \nonumber\\
&&F_{\mu\nu}^2=b^2\, C_{\mu\nu}^2 +a^2\, B_{\mu\nu}^2, \quad
\eta_{\mu\nu} =C_{\mu\nu}^2 -B_{\mu\nu}^2. \label{CBinv}
\end{eqnarray}
Most importantly, we get the {\em decomposition relations} by a
straightforward calculation
\begin{eqnarray}
&&(CB)_{\mu\nu}=0=(BC)_{\mu\nu}, \quad C_{\mu\nu}^3=C_{\mu\nu}, \quad
B^3_{\mu\nu} =-B_{\mu\nu}, \nonumber\\
&&C^{2\mu}{}_\mu\equiv -C_{\mu\nu}C^{\mu\nu} =-2, \quad
B^{2\mu}{}_\mu\equiv -B_{\mu\nu}B^{\mu\nu} =2,
\end{eqnarray}
where it becomes obvious that the decomposition of the field strength
tensor into $C$ and $B$ corresponds to a decomposition into orthogonal
magnetic and electric subspaces. 

Employing this representation of the field strength tensor, we can
decompose any function of $F_{\mu\nu}$, regular at $F=0$, into
\begin{eqnarray}
f(F)_{\mu\nu} &=& f(-bC)_{\mu\nu} + f(aB)_{\mu\nu} \nonumber\\
&=& f_{\text{odd}}(-b)\, C_{\mu\nu} + f_{\text{even}}(-b) C^2_{\mu\nu} 
-\I\, f_{\text{odd}}(\I a)\, B_{\mu\nu} -f_{\text{even}}(\I a)\,
B^2_{\mu\nu}, \label{CBdecomp}
\end{eqnarray}
where $f_{\text{even,odd}}$ denotes the even/odd part of $f$.

\subsection*{Decomposition of $\Pi_{55}$}
The axialvector--axialvector amplitude in an arbitrary external field
is defined by
\footnote{Our Dirac matrix conventions are 
$\lbrace\gamma_{\mu},\gamma_{\nu}\rbrace
= -2\eta_{\mu\nu}\mathbbm{1}$, $\gamma_5^2 =1$.} 
\begin{equation}
\Pi_{55}^{\mu\nu}(k)=\I e_5^2 \tr_\gamma\, \int_p \gamma^\mu \gamma_5
\, g(p) \, \gamma^\nu \gamma_5\, g(p-k), \label{Pi1}
\end{equation} 
where $g(p)$ denotes the Fourier transform of the Dirac Green's
function, and $\int_p=\int \frac{d^4p}{(2\pi)^4}$. In constant
external fields, this object is given by 
\cite{Dittrich:2000zu}
\begin{equation}
g(p)=\I\int\limits_0^\infty ds\, \E^{-\I m^2s}\E^{-Y(\I s)} \Bigl[ m
-\gamma_\alpha \bigl( p-\I e\mathsf{FX}p\bigr)^\alpha \Bigr] \E^{-
  p\mathsf{X} p} \E^{\frac{\I e}{2} s\sigma F}, \label{11124}
\end{equation}
where
\begin{equation}
Y(s)          = \frac{1}{2} \tr \ln [ \cos (e\mathsf{F} s)], \quad
\mathsf{X}(s) = \frac{\tan (e\mathsf{F}s)}{e\mathsf{F}},
\end{equation}
and we employed matrix notation, $\sigma \mathsf{F} =\sigma_{\mu\nu}
F^{\mu\nu}$, $\sigma_{\mu\nu}=(\I/2) [\gamma_\mu,\gamma_\nu]$. Using
\begin{equation}
[\E^{\I \frac{e}{2} s\sigma \mathsf{F}},\gamma_5]=0, \label{commut}
\end{equation} 
it can be shown that
\begin{equation}
g(p)\, \gamma_5 = -\gamma_5 \, g(p)\big|_{m\to -m}\bigr( =\gamma_5\,
g(-p)\bigl). \label{commut2}
\end{equation}
Adding and subtracting the mass term with the correct sign, we get
\begin{equation}
g(p)\big|_{m\to -m} =g(p) -2\I  \, m
\int\limits_0^\infty ds\, \E^{-\I m^2s}
\E^{-Y(\I s)}\, \E^{-p\mathsf{X}p} \E^{\I\frac{e}{2} s\sigma
  \mathsf{F}} .\label{gminusm}
\end{equation}
Inserting Eqs.\re{gminusm} and\re{commut2} into Eq.\re{Pi1}, we again
find a decomposition\footnote{Our definition of $\Pi_{\rm spin}^\mn$
here differs by a sign from the one used in \cite{Dittrich:2000zu} (it
agrees with \cite{weinberg}).}

\begin{equation}
\Pi_{55}^{\mu\nu}(k)=
{e_5^2\over e^2}
\Pi^{\mu\nu}_{\text{spin}} (k) +N_{55}^{\mu\nu}(k),
\label{decompPi}
\end{equation}
where $\Pi_{\rm spin}^{\mu\nu}$ denotes the vector
polarisation tensor, and the additional term arising from the second
term of Eq.\re{gminusm} is defined by
\begin{eqnarray}
&&N_{55}^{\mu\nu}(k)=2\I e_5^2 m^2 \int_p \int\limits_0^\infty ds
ds'\, \E^{-\I m^2(s+s')} \E^{-(\mathsf{Y}'+\mathsf{Y})}
\E^{-p(\mathsf{X}+\mathsf{X}')p} \E^{2p\mathsf{X}' k}
\E^{-k\mathsf{X}'k}\nonumber\\
&&\qquad\qquad\qquad\qquad\qquad\qquad\times
 \tr_\gamma\, \bigl[ \gamma^\mu \E^{\I \frac{e}{2}
  s\sigma \mathsf{F}}  \gamma^\nu \E^{\I \frac{e}{2}
  s'\sigma \mathsf{F}} \bigr]. \label{N1}
\end{eqnarray}

\subsection*{Calculation of $N_{55}^{\mu\nu}$}

The $p$ integral is Gaussian and gives (in Minkowski space)
\begin{equation}
\int_p \E^{-p(\mathsf{X}+\mathsf{X}')p}
\E^{2p\mathsf{X}' k} \E^{-k\mathsf{X}'k} 
= \frac{\I}{(4\pi)^2} \, \bigl( \det (\mathsf{X}+\mathsf{X}')
\bigr)^{-1/2} \, \E^{-k
  \frac{\mathsf{X}\mathsf{X}'}{\mathsf{X}+\mathsf{X}'}
  k}. \label{pint}
\end{equation}
Inserting Eq.\re{pint} into Eq.\re{N1}, we encounter the combination
\begin{equation}
\E^{-(\mathsf{Y}'+\mathsf{Y})}\, \bigl( \det (\mathsf{X}+\mathsf{X}')
\bigr)^{-1/2}= \sqrt{ \det \frac{ e\Fsf}{\sin e\Fsf \I \ts}}, \quad
\text{where}\,\, \ts:= s+s'. \label{pint2}
\end{equation}
Putting it all together, the desired quantity reads
\begin{eqnarray}
N_{55}^{\mu\nu}(k)&=&-\frac{2 e_5^2 m^2}{(4\pi)^2} \int\limits_0^\infty ds
ds'\, \E^{-\I m^2\ts} \, \sqrt{ \det \frac{ e\Fsf}{\sin e\Fsf \I \ts}} 
\E^{-k \frac{\mathsf{X}\mathsf{X}'}{\mathsf{X}+\mathsf{X}'}  k}
\, \tr_\gamma\, \bigl[ \gamma^\mu \E^{\I \frac{e}{2}
  s\sigma \mathsf{F}}  \gamma^\nu \E^{\I \frac{e}{2}
  s'\sigma \mathsf{F}} \bigr].
\non\\ \label{N2}
\end{eqnarray}
Employing the field strength tensor decomposition described above,
it is straightforward to show that
\begin{equation}
\sqrt{ \det \frac{ e\Fsf}{\sin e\Fsf \I \ts}} =-\frac{ea\,eb}{\sinh
  eb\ts\, \sin ea\ts}, \label{sins}
\end{equation}
and that the exponent in Eq.\re{N2} yields
\begin{eqnarray}
k \frac{\mathsf{X}\mathsf{X}'}{\mathsf{X}\!+\!\mathsf{X}'}  k
&=& -\I \left[
  \frac{\cosh ebv \ts -\cosh eb\ts}{2eb\,\sinh eb\ts} \,
  kC^2k 
  +  \frac{\cos eav \ts -\cos ea\ts}{2ea\,\sin ea\ts}\,
  kB^2k\right]\!,\,\, 
\non\\ \label{expo}
\end{eqnarray} 
where $v=1-\frac{2s'}{\ts}$.
Read together with the mass term $\E^{-\I m^2 \ts}$ in Eq.\re{N2}, we
find the usual phase factor of the polarisation tensor or the
axialvector--vector amplitude
\begin{equation}
\E^{-\I m^2 \ts}\,
\E^{-k \frac{\mathsf{X}\mathsf{X}'}{\mathsf{X}+\mathsf{X}'}  k}
=\E^{-\I \ts \phi_0}, \label{expo2}
\end{equation}
where 
\begin{equation}
\phi_0 = m^2 -\frac{k C^2 k}{2} \frac{\cosh ebs -\cosh v ebs}{ebs
           \sinh ebs}
             -\frac{k B^2 k}{2} \frac{\cos v eas -\cos eas}{eas
           \sin eas}.
\end{equation}
The final task is to find an appropriate representation of the Dirac
trace in Eq.\re{N2}. The trace has been evaluated, for instance, in
App.~C of \cite{Dittrich:2000zu}; in terms of the $C$ and $B$
decomposition the result reads
\begin{eqnarray}
\tr_\gamma\, \bigl[ \gamma_\mu \E^{\I \frac{e}{2}
  s\sigma \mathsf{F}}  \gamma_\nu \E^{\I \frac{e}{2}
  s'\sigma \mathsf{F}} \bigr] 
&=&4\biggl\{ -\cos ea\ts\cosh ebv\ts\, C_{\mu\nu}^2 
  +\cos eav\ts \cosh eb\ts\, B^2_{\mu\nu} \nonumber\\
&&\qquad +\cos ea\ts \sinh ebv\ts \, C_{\mu\nu}
  -\sin eav\ts \cosh eb\ts\, B_{\mu\nu} \biggr\}.  
\non\\ \label{trace2}
\end{eqnarray}
Upon substitution of $\int_0^\infty d\ts\, \ts\int_{-1}^1
\frac{dv}{2}$ for $\int_0^\infty ds ds'$, we observe that the $v$
integration is even, the phase factor is even in $v$, but the terms
$\sim C_{\mu\nu},B_{\mu\nu}$ are odd and thus drop out. The final
result before renormalisation then reads ($\ts\to s$)
\begin{eqnarray}
N_{55}^{\mu\nu}(k)=\frac{e_5^2}{2\pi^2} m^2 \int\limits_0^\infty
\frac{ds}{s} \int\limits_{-1}^1 \frac{dv}{2}\!\!\! &&\!\!\!\E^{-\I
  s\phi_0}\, \frac{eas\, ebs}{\sin eas \sinh ebs}  \label{N3}\\
&&\times\bigl[ -\cos eas \cosh ebv s\,
C_{\mu\nu}^2 +\cos eav s \cosh ebs \, B_{\mu\nu}^2
\bigr],\nonumber
\end{eqnarray}
in agreement with (\ref{vp55final}) since 
$B_\mn^2 = (\hat{\cal Z}_+^2)_\mn,
C_\mn^2 = - (\hat{\cal Z}_-^2)_\mn$.

\section{Discussion}
\renewcommand{\theequation}{5.\arabic{equation}}
\setcounter{equation}{0}

We have used both the worldline formalism and
standard Feynman diagrams for a first calculation
of the axialvector vacuum polarisation tensor
in an arbitrary constant electromagnetic field.
With both methods, we obtained 
the same compact integral representation
for the difference between this tensor and the
well-known vector vacuum polarisation tensor in a 
constant field.
For the special case of a purely magnetic field,
our result agrees with the one of \cite{kuzmik};
the case of a general constant field has,
as far as is known to the authors, not been
treated before.

As in the vector and vector--axialvector cases, 
the performance of suitable partial integrations
has turned out to be essential for reaching a
maximally compact integral representation in the
worldline formalism. However, while in those cases the
appropriate partial integrations involved only the
loop variables, and could be easily
found following the Bern-Kosower prescription
\cite{berkos} of removing all second derivatives
of worldline Green's functions, in the present
case a judiciously chosen combination of partial
integrations in both the loop variables and the
global proper-time was found necessary for this purpose.
It would be clearly useful to investigate the systematics
of this partial integration procedure for the
general vector--axialvector amplitudes, starting from
the (vacuum) master formulas given in \cite{dimcsc}. 
 
\vskip15pt
\noindent{\bf Acknowledgements:}
We would like to thank S.L.~Adler, T.~Binoth, W.~Dittrich and
M.~Kachelriess for various helpful informations.

\end{document}